# COMPUTING HIGHEST DENSITY REGIONS FOR CONTINUOUS UNIVARIATE DISTRIBUTIONS WITH KNOWN PROBABILITY FUNCTIONS

BEN O'NEILL[*], *Australian National University*[**]

WRITTEN 16 JANUARY 2020; REVISED 16 JUNE 2021

**Abstract**

We examine the problem of computing the highest density region (HDR) in a computational context where the user has access to a density function and quantile function for the distribution (e.g., in the statistical language `R`). We examine several common classes of continuous univariate distributions based on the shape of the density function; this includes monotone densities, quasi-concave and quasi-convex densities, and general multimodal densities. In each case we show how the user can compute the HDR from the quantile and density functions by framing the problem as a nonlinear optimisation problem. We implement these methods in `R` to obtain general functions to compute HDRs for classes of distributions, and for commonly used families of distributions. We compare our method to existing `R` packages for computing HDRs and we show that our method performs favourably in terms of both accuracy and average speed.

HIGHEST DENSITY REGION; INTENSITY FUNCTION; UNIVARIATE DISTRIBUTION; MONOTONICITY; QUASI-CONCAVITY; NONLINEAR OPTIMISATION.

## 1. Introduction

In various statistical contexts, it is useful to summarise a distribution by finding a region on its support with a specified minimal coverage probability. This task may be useful in descriptive analysis or when making predictions from the distribution. In both cases, it is common to want to find the *smallest* region with a specified minimum coverage probability, to be as accurate as possible about the probable location of a random variable generated from that distribution. For a wide range of continuous distributions, it is known that the smallest region with a specified coverage probability is obtained by taking all points in the support with some minimum density (see e.g., Box and Tiao 1973), which leads to the notion of the "highest density region" (HDR).

Since HDRs encompass all points satisfying some minimal density cut-off, if the density is continuous then the boundaries of the HDR are the "level sets" of the density at a stipulated value (computed or estimated from the specified coverage probability). Thus, the mathematical task of finding the HDR is similar to finding level sets of a function — specifically, one wishes to find the level sets of the density corresponding to some stipulated coverage probability for the corresponding HDR. Hyndman (1996) gives a general treatment of the task of computing

---

and graphing HDRs for univariate and multivariate distributions, and he notes that there are two broad approaches. HDRs can be computed analytically using the "numerical integration approach", which requires integration to obtain either the quantile function of the distribution, or equivalent information that allows the user to form intervals with given levels of coverage. Alternatively, in order to avoid difficulties with numerical integration, HDRs can also be computed using the "density quantile approach" where one uses sampling methods to estimate the density cut-off for the HDR, and then uses grid methods to search for the level sets for this cut-off value over the support of the distribution.

Existing statistical literature on HDRs mostly look at general cases, allowing for multivariate distributions and multimodality. In particular, Hyndman (1996) and Kruschke (2015) show the density quantile approach to create broad methods applicable to multivariate distributions, and can allow for distributions with multiple modes. This general treatment makes heavy use of approximation methods involving grid searches, owing to the highly general nature of the functions under analysis. In the present paper, our focus is much narrower, and our ambitions are more modest: to formulate a set of optimisation procedures to compute the HDR in contexts where the user already has available probability functions (e.g., quantile function, distribution function, density function, etc.) for a univariate unimodal distribution. Our analysis will look at classes of densities based on their shape, and we will cover cases that encompass all the most common families of univariate distributions. We will also show how this can be extended to more general multimodal densities that do not have simple shape properties. Our approach employs analytic optimisation techniques that yield exact HDRs — by restricting attention to classes of densities with known shape properties, we do not require approximations such as grid searches or Monte-Carlo estimation of the density cut-off for the HDR. Consequently, we are able to create algorithms to compute exact HDRs over wide classes of densities, including common families of distributions used in statistical analysis.

To further motivate the scope of our analysis, it is worth saying a word about the coverage of statistical packages for dealing with standard families of univariate probability distributions. The base package of the statistical language `R` has probability functions for a range of common distributional families, including quantile and density functions,[1] but it does not have functions for computing HDRs from these distributions. It is desirable for users to have the capacity to compute HDRs accurately at least over these base distributions, and also over broad families of distributions in extension packages. These distributional families have shape properties that make them amenable to methods designed for distributions of particular shapes, and in this paper we pursue this method to create HDR functions for all the base distributions in `R`.

---

[1] These distributions are in the `stats` package, which is now included in the base version of `R`.

Here is it worth noting that there are existing packages written by other authors —which we will discuss in detail below— that give functions to compute HDRs for general density inputs. Unfortunately, the generality of the inputs used in these functions means that each package has major drawbacks in computing HDRs over the base distributions in `R`. Broadly speaking, the existing packages suffer from one of two problems: either they use approximation methods to give highly general —but only approximate— computation of the HDR; or they give "exact" computations that make assumptions about the shape of the distribution that do not hold over all the base distributions in `R`. In the latter case, the "exact" calculations work for some classes of distributions but give incorrect results for other classes of distributions; the user receives no warning that the method is inaccurate when used on a non-compliant density.

The methods set out in this paper are programmed in the `stat.extend` package (O'Neill and Fultz 2020). In a later section of the paper we will compare our method to other existing packages that compute HDRs, and we will show that none of the other packages succeeds in accurate (non-approximate) computation of HDRs over all base distributions in `R`. Our own method and its corresponding package succeeds in giving exact HDRs for all base distributions in `R` and several distributions in extension packages. By focussing on individual distributional families based on the shape of the density, our method is less general, but more accurate, than existing `R` packages.

**2. HDRs and intensity functions**

Before proceeding to our main analysis, we will first introduce some basic theory to motivate the use of HDRs, and to clarify the distinction (in some cases) between the "minimum coverage probability" and the "actual coverage probability" of a HDR. We also introduce a function that we will call the "intensity function", which relates closely to the HDR. The main reason that HDRs are of interest in statistical applications is that they correspond to the smallest region with a stipulated minimum coverage probability. To establish this, we will present some basic definitions and theorems where we consider a real scalar random variable $X$ with density $f$. To simplify our analysis, we restrict our attention to closed sets, so we will define the HDR as a closed set, and we look at smallest regions over all closed sets; to measure "smallness" we denote the measure of any set $\mathcal{A}$ by $|\mathcal{A}|$.[2]

**DEFINITION (Smallest closed covering region):** For any value $0 \leq \alpha \leq 1$ a "smallest closed covering region" with minimum coverage probability $1 - \alpha$ is any closed set $\mathcal{H}$ that satisfies the following conditions:

(Coverage) $\quad\quad\quad \mathbb{P}(\mathcal{H}) \geq 1 - \alpha,$

(Minimisation) $\quad\quad |\mathcal{H}| \leq |\mathcal{A}|$ for all closed $\mathcal{A}$ with $\mathbb{P}(\mathcal{A}) \geq 1 - \alpha.$

**DEFINITION (Intensity function):** For any random variable $X$ with density function $f$, the corresponding intensity function $H: \mathbb{R} \to [0,1]$ is defined as:

$$H(a) \equiv \mathbb{P}(f(X) \geq a) \quad\quad \text{for all } a \in \mathbb{R}.$$

This is a non-increasing function with $H(0) = 1$ and $\lim_{a \to \infty} H(a) = 0.$[3]

**DEFINITION (Highest density region):** For any random variable $X$ and any value $0 \leq \alpha \leq 1$ we define the "highest density region" with minimal coverage probability $1 - \alpha$ to be the set:

$$\mathcal{H} \equiv \text{closure}\{x \in \mathbb{R} | f(x) \geq f_*\} \quad\quad f_* \equiv \inf\{a \in \mathbb{R} | H(a) \geq 1 - \alpha\}.$$

The value $1 - \alpha$ is the "minimum coverage probability" and the value $\mathbb{P}(X \in \mathcal{H})$ is the "actual coverage probability" of the region (the latter must be at least as large as the former).

The above definition establishes the concept of a HDR for a scalar random variable. The HDR can be regarded as a useful generalisation of the "mode" of the distribution, insofar as every non-empty HDR contains points in the mode (and only contains points not in the mode if it has already included all points in the mode).[4] As $\alpha$ decreases, the HDR expands its coverage until it eventually becomes the closure of the whole support. Even if there is not a unique mode (i.e., the mode is a set of points instead of a unique point), the HDR will still start by adding this set of points before adding any other points.

---

[2] The notion of the "smallest" region requires us to work in a probability space with sufficient structure that we can measure the size of an event in the sample space. Our probability space takes the set of real numbers as the sample space and the corresponding class of Borel sets as the sigma-field of all measureable events. Within this context, every possible "region" under analysis is a Borel set. For discrete random variables the "smallness" of a region is measured by counting measure (i.e., the number of points in the region), and for absolutely continuous random variables it is measured by Lebesgue measure.

[3] These properties are trivial to establish. Since $f$ is a density function, it is non-negative, so we have $H(a) = 1$ for all $a \leq 0$. Since $a < a'$ implies $\{f(X) \geq a\} \supseteq \{f(X) \geq a'\}$ we also have $\mathbb{P}(f(X) \geq a) \geq \mathbb{P}(f(X) \geq a')$ so $H$ is non-increasing. Note that we can restrict the domain of $H$ to $\mathbb{R}_{0+}$ without loss of usefulness. Here we have allowed the domain to include the whole real line, but this choice makes no difference to our subsequent analysis.

[4] More specifically, if the set of points in the mode has zero probability, then every non-empty HDR will contain the entire set of points in the mode. Contrarily, if the set of points in the mode has zero probability, then every non-empty HDR will contain some of these points, up to the required coverage probability; and will only contain points outside the mode if it has not yet met the required coverage probability.

**LEMMA 1:** If $\mathcal{H}$ is a highest density region then for any set $\mathcal{A}$ we have:

$$|\mathcal{A}| < |\mathcal{H}| \quad \Longrightarrow \quad \mathbb{P}(X \in \mathcal{A}) < \mathbb{P}(X \in \mathcal{H}),$$
$$|\mathcal{A}| \leq |\mathcal{H}| \quad \Longrightarrow \quad \mathbb{P}(X \in \mathcal{A}) \leq \mathbb{P}(X \in \mathcal{H}).$$

**THEOREM 1:** A highest density region $\mathcal{H}$ with actual coverage probability $1 - \alpha$ is a smallest closed covering region with minimal coverage probability $1 - \alpha$.

**THEOREM 2:** If the intensity function for $X$ is continuous, then a highest density region $\mathcal{H}$ formed with minimal coverage probability $1 - \alpha$ also has actual coverage probability $1 - \alpha$.

The above theorems establish that the HDR is a smallest closed covering region, and also shows the relationship between the minimum coverage probability and actual coverage probability. The actual coverage probability is equal to the minimum coverage probability if the intensity function is continuous, but it may be higher than the minimum coverage probability if the intensity function $H$ is discontinuous. This latter case will occur in two situations: (1) if $X$ is discrete then the intensity function is a step function, with jumps at each value corresponding to the probability of an outcome in the support, and so the intensity function is also a step function; (2) if $X$ is continuous but has one or more neighbourhoods on its density that are flat (with non-zero density) then the intensity function will jump up at those density values. In either situation the discontinuity of the intensity function causes the actual coverage probability to be higher than the stipulated minimum coverage probability for some inputs.

We examine the computation of the highest density region (HDR) and the intensity function where we have a real scalar random variable with distribution function $F$, quantile function $Q$, density function $f$ and logarithmic-derivative-density $u$.[5] Our focus is on continuous random variables where the density function is differentiable over the support (i.e., where $u$ exists over the support). In our analysis we will assume that these probability functions are all available, and we will frame our analysis in terms of nonlinear optimisation. In our later implementation in `R`, we will see that not all of these probability functions are needed; the user can actually use numerical differentiation methods without the density or the logarithmic-derivative-density if needed. We split our analysis into four cases, entailing increasing complexity.

---
[5] The function $u: \mathbb{R} \to \mathbb{R}$ is defined by $u(x) = f'(x)/f(x)$ for all $x \in \mathbb{R}$. It can be obtained as the derivative of the log-density function. (By convention we set $u(x) = 0$ when $f'(x) = f(x) = 0$ and we set $u(x) = \infty$ when $f'(x) > 0$ and $f(x) = 0$.)

| **Densities for Continuous Random Variables** ||
| --- | --- |
| **Class of density** | **Common distributions in this class** |
| **Monotone density** 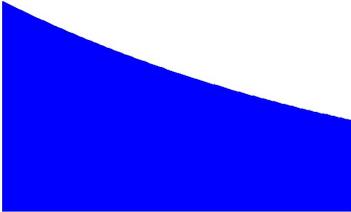 | - **Exponential** distribution; <br> - **Gamma** and **Weibull** distributions (with shape parameter no greater than one); <br> - **Beta** distribution (with one shape parameter no less than one and the other no greater than one); <br> - **F** distribution (with numerator degrees-of-freedom no greater than two). |
| **Quasi-concave density** 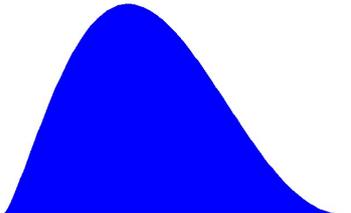 | - **Normal** and **lognormal** distributions; <br> - **Student T** distribution; <br> - **Gamma** and **Weibull** distributions (with shape parameter greater than one); <br> - **Beta** distribution (both shape parameters greater than one) <br> - **F** distribution (with numerator degrees-of-freedom greater than two). |
| **Quasi-convex density** 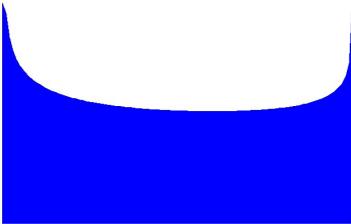 | - **Beta** distribution (both shape parameters less than one). |
| **Other densities** 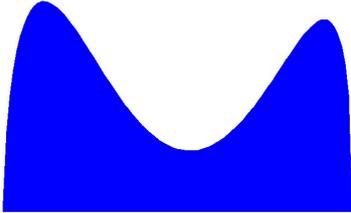 | None of the families of distributions we examine here fall into this case, but we will discuss it for completeness. |

## 3. Computing HDRs — the "up-down" method and the "left-right" method

Before introducing specific methods for computing HDRs over various classes of continuous univariate distributions, it is worth differentiating two general classes of iterative optimisation methods for this task, which we will call the "up-down" method and the "left-right" method. This is simplest to introduce in the context of a simple unimodal continuous distribution, where the HDR is a single closed interval, so we will assume that context for the present explanation. In the **up-down method** we use the density cut-off value as the optimisation variable, and we seek to minimise the distance between the actual coverage probability of the interval and the stipulated minimum coverage probability $1 - \alpha$. At each iteration of the optimisation routine we compute the interval containing points with density values at least as large as this minimum density cut-off (in our argument variable), then we compute the coverage probability for that interval, and then we move the cut-off density up or down to obtain the appropriate coverage probability. Contrarily, in the **left-right method** we use the position of one of the boundaries of the interval (say, the lower boundary) as the optimisation variable and we compute the other boundary using the stipulated minimum coverage probability for the region, and then we move the argument boundary left or right until we obtain the smallest possible interval width. These methods are summarised in the table below.

|  | Optimisation problems for computing HDR (considered here for unimodal continuous densities) | |
|---|---|---|
|  | **Up-down method** | **Left-right method** |
| **Optimisation**[6] | Minimise $[H(a) - (1-\alpha)]^2$ | Minimise $U(L, \alpha) - L$ |
| **Argument** | Density cut-off $a \geq 0$ | Lower interval boundary $L$ |
| **Iterations** | (1) Compute the region $\mathcal{H}(a)$ corresponding to the density cut-off (usually using grid methods); <br> (2) Compute intensity $H(a)$ at the density cut-off (usually using grid methods); <br> (3) Shift cut-off $a$ up or down to change intensity. | (1) Compute the upper interval boundary $U(L, \alpha)$ that gives coverage probability $1 - \alpha$ (using the quantile function); <br> (2) Compute interval width; <br> (3) Shift lower interval boundary $L$ left or right to change the interval width. |
| **Fixed by method** | Highest-density property is fixed by step (1) of the iteration method | Coverage probability is fixed by step (1) of the iteration method |

---

[6] The up-down method is essentially a root-finding problem where we wish to solve $H(a) = 1 - \alpha$ for $a$, but this can easily be framed as an optimisation problem, to give the benefits of derivatives, etc., for rapid convergence to the solution. We frame the method as an optimisation here to give a clearer comparison to the left-right method.

The up-down method can be summarised by saying that it fixes the interval to one with highest density and then *optimises the coverage probability*. The left-right method can be summarised by saying that it fixes the coverage probability and then *optimises the interval width*. Each of these methods of posing the optimisation has its own advantages and disadvantages. The main advantage of the up-down method is that it is extremely general; it can be used in combination with grid-methods to compute approximate HDRs in multivariate problems, or in univariate problems where the density has a complex shape (e.g., several local maximums). By going directly to optimisation of the density cut-off, this method does not require knowledge of the "shape" of the distribution under consideration. The main disadvantage is that the use of grid-methods adds an additional approximation to the method, since it essentially approximates a continuous density function by a step-function over a fine grid. Consequently, the method is highly general, and relatively simple to apply, but it has approximating elements that lead to some computational error. The up-down method is the method used in Hyndman (1996) and Kruschke (2015) and is implemented in general packages like **`hdrcde`**.

The left-right method is much less general than the up-down method, but it has advantages in some problems. Since it optimises using boundaries of the HDR, the method requires enough preliminary knowledge of the "shape" of the distribution to specify the number and type of boundaries of the HDR in advance (e.g., in the continuous unimodal case we know in advance that the HDR is a single closed interval). This is possible for most univariate problems and even some multivariate HDR problems,[7] but it become increasingly complex as the dimensions of the distribution increase, or as the shape becomes more complex. Consequently, the left-right method is not suitable for posing algorithms that apply generally to distributions of any possible shape or dimension. Notwithstanding this lack of generality, the method has some advantages. Because it involves preliminary knowledge of the "shape" of the distribution, it does not require grid-methods, and so it does not suffer from the consequent computation error that comes from employing step-function approximations to a continuous density. Because it uses shape information, it is also able to avoid situations where remote areas of the density function are missed by grid-search methods. Consequently, when employed in the proper context, it tends to be more accurate but less general.

---

[7] This can be done in cases where the HDR for a multivariate distribution can be reduced to a corresponding optimisation problem for a univariate distribution. For example, the HDR for a multivariate normal distribution is a closed set bounded by the ellipsoid using the Mahalanobis distance. The ellipsoid can be computed using an optimisation for the univariate chi-squared distribution.

The up-down optimisation method is generally less accurate due to the use of grid methods, but both methods suffer from arithmetic error and optimisation tolerance (i.e., the fact that the iteration terminates in a finite number of steps, subject to some tolerance level). While the two methods have the same mathematical solution in theory, in practice there are discrepancies in outputs due to different approximations used internally in the method, and different ways that arithmetic error and optimisation tolerance enter into the methods. In the up-down method the "highest-density" condition is fixed by the method, but the coverage probability of the interval is subject to arithmetic error and error due to optimisation tolerance, as well as approximation error from grid methods. Consequently, the HDR computed by the up-down method should may have a coverage probability that is slightly different to the stipulated minimum coverage probability (it may even be slightly *less* than the stipulated minimum[8]). In the left-right method the coverage probability of the region is fixed by the method, but the boundaries of the interval are subject to arithmetic error and error due to optimisation tolerance. Consequently, the HDR computed by the left-right method will have the correct coverage probability, but the density values at its boundary points may be slightly different.

In the present paper we set out specific optimisation methods using the "left-right" method for a range of different "shapes" for the continuous univariate distribution under examination. We will examine optimisation procedures for monotone and unimodal (quasi-concave) densities, bimodal densities that are quasi-convex, and more general classes of densities that are neither quasi-concave or quasi-convex, and have multiple local maxima and at least one local minima. These broad classes of distribution shapes can be used to create HDR computation procedures applying to "families" of continuous distributions, by first determining the relevant shape of the density from the parameters, and then applying the requisite method for that shape. HDR functions using these methods are implemented for all standard families of distributions in `R` in the `stat.extend` package, which we examine at the end of the paper. We compare the accuracy and speed of this package with other packages that use the "up-down" method for HDR computation, and we find that our method gives superior results.

---

[8] In most statistical applications, it is not usually problematic if the computed HDR has a coverage probability that is slightly below the stipulated minimum. If this minimum bound is important then it is possible to change the optimisation method to disallow outcomes that fail to meet the stipulated minimum coverage probability. This can be done either using "penalty" methods or by using iterative procedures that end by going back over the iterations to search for the last iteration that met the required minimum bound requirement.

## 4. HDRs for monotone or quasi-concave densities

Our analysis will be for a continuous distribution with a differentiable density function $f$. To frame our problem, we first note that, for a continuous distribution, any closed interval $[L, U]$ with a fixed coverage probability can be fully characterised by the parameter $\theta \equiv F(L)$, which is the probability of an outcome in the lower tail, below the interval.[9] If the interval has some fixed coverage probability $1 - \alpha$ then this parameter is in the range $0 \leq \theta \leq \alpha$. Since the distribution is assumed to be continuous, the upper and lower bounds of the interval can be written as $L = Q(\theta)$ and $U = Q(\theta + 1 - \alpha)$, so the width of the interval is:

$$W(\theta) = Q(\theta + 1 - \alpha) - Q(\theta) \qquad \text{for all } 0 \leq \theta \leq \alpha.$$

The case of a **monotone density function** is trivial. If the density is monotone decreasing (or non-increasing) then taking $\theta = 0$ gives the HDR with bounds $L = Q(0)$ and $U = Q(1 - \alpha)$, and $f_* = f(U)$ is the density cut-off for the HDR. If the density is monotone increasing (or non-decreasing) then taking $\theta = \alpha$ gives the HDR with bounds $L = Q(\alpha)$ and $U = Q(1)$, and $f_* = f(L)$ is the density cut-off for the HDR. The special case of a uniform density is both non-increasing and non-decreasing, so the HDR is any interval with appropriate coverage that is fully concentrated on the closure of the support. (In this case it is conventional to set the HDR to be the "middle" interval in the support.) In any of these cases, it is sufficient to have the quantile function $Q$ to compute the HDR.

The more general case of a **quasi-concave density** function is non-trivial, and requires us to solve an optimisation problem. Since the coverage probability is fixed by the stipulation of the required minimum coverage probability for the HDR, we can find the HDR by minimising the objective function $W$ over the constrained range of the parameter $\theta$. We then substitute the optimising value of this parameter to obtain the optimised lower and upper bounds of the HDR. We will assume that the density function is strictly quasi-concave (i.e., strictly unimodal)[10] and we will also assume that $0 < \alpha < 1$ to remove trivial cases. (The cases where $\alpha = 0$ or $\alpha = 1$ can be dealt with directly by stipulation; in the first case the HDR is the empty sets and in the second case the HDR is the closure of the support.)

---

[9] Since the random variable $X$ is assumed to be continuous, we have $\mathbb{P}(X < L) = \mathbb{P}(X \leq L) = F(L) = \theta$, so the parameter is the probability of the random variable being below a closed interval bounded below at $L$.

[10] The case of the uniform distribution, which is not strictly unimodal, will be considered separately at the end of our analysis. This will ensure that our method covers all continuous unimodal distributions in base R.

This optimisation problem can be solved using ordinary calculus techniques. Since $Q = F^{-1}$ we first apply the inverse function theorem to obtain the useful derivative rule:

$$\frac{d}{d\theta} Q(g(\theta)) = \frac{g'(\theta)}{f(Q(g(\theta)))}.$$

Taking $u(x) = f'(x)/f(x)$ to be the derivative of the logarithm of the density function,[11] and using the above derivative rule, the objective function has first and second derivatives:

$$W'(\theta) = \frac{1}{f(U(\theta))} - \frac{1}{f(L(\theta))},$$

$$W''(\theta) = \frac{u(L(\theta))}{f(L(\theta))^2} - \frac{u(U(\theta))}{f(U(\theta))^2}.$$

Since the density function $f$ is assumed to be differentiable and strictly quasi-concave, it has a unique mode $\hat{x}$ and its slope satisfies the equation $\operatorname{sgn} u(x) = \operatorname{sgn}(\hat{x} - x)$ for all $x \in \mathbb{R}$. Thus, for any parameter value $\theta$ giving bounds $L(\theta) < \hat{x} < U(\theta)$ that span the mode, we must have $u(L(\theta)) > 0$ and $u(U(\theta)) < 0$ which implies that $W''(\theta) > 0$. We will refer to this range of parameter values where the bounds span the mode as the "admissible range" (see below). We therefore see that the width function is strictly convex over the admissible range of parameter values $\theta$.

**THE "ADMISSIBLE RANGE" FOR STRICTLY QUASI-CONCAVE DENSITY:** When dealing with a strictly quasi-concave function that is not monotonic, the bounds of the HDR must always span the mode, which gives the inequality $L(\theta) < \hat{x} < U(\theta)$ defining the admissible range. For any value $0 < \alpha < 1$ the admissible range of $\theta$ is the range:

$$\max(0, F(\hat{x}) - (1-\alpha)) < \theta < \min(\alpha, F(\hat{x})).$$

(This result is obtained by inverting the pair of inequalities defining the admissible range, and overlaying the result onto the additional requirement that $0 \leq \theta \leq \alpha$.) In the sections outside the admissible range the width function is monotonic, but it is not necessarily convex. It is also worth noting that if $\alpha \leq F(\hat{x}) \leq 1 - \alpha$ then the admissible range is the full range $0 \leq \theta \leq \alpha$. So long as the mode is not too close to one end of the support, and so long as $\alpha$ is not too small, this latter inequality will be satisfied. In most HDR calculations of interest in practice, the admissible range is the full range. □

---

[11] Note that the function $u$ should not be confused with the score function for the random variable. The latter is the derivative of the logarithm of the *likelihood function*, with respect to the *parameters* of the distribution. The function $u$ is instead the derivative of the logarithm of the density, with respect to the *outcome value*.

The above results show that the width function is strictly convex over the "admissible range" of parameter values that lead to bounds spanning the mode of the density. The optimisation problem therefore has unique critical point $\hat{\theta}$ that is the minimising parameter, and this gives HDR bounds $\hat{L} < \hat{x} < \hat{U}$ that span the mode $\hat{x}$ of the distribution. The bounds are characterised by the first-order condition $f_* = f(\hat{L}) = f(\hat{U})$, which is the well-known intuitive condition for a HDR. Since the width function is strictly convex over the admissible range, the second order condition for a local minimum is also satisfied. Solution to the critical point equation may be analytically possible for some density functions, but it will generally require iterative methods. A standard method is Newton-Raphson iteration using the recursive equation:

$$\hat{\theta}_{k+1} = \hat{\theta}_k - \frac{W'(\hat{\theta}_k)}{W''(\hat{\theta}_k)} = \hat{\theta}_k - \frac{f(\hat{L}_k)f(\hat{U}_k)(f(\hat{L}_k) - f(\hat{U}_k))}{u(\hat{L}_k)f(\hat{U}_k)^2 - u(\hat{U}_k)f(\hat{L}_k)^2}.$$

Taking any starting value $\hat{\theta}_0$ in the admissible range of the parameter (i.e., giving bounds that span the mode) ensures that the first iteration of the Newton-Raphson algorithm is in the correct direction. If the mode of the distribution is known, it should be possible to choose a starting parameter that is known to be in the admissible range, but even if one starts outside this range, convergence to the optima will occur for most densities. It is also worth noting that the starting value $\hat{\theta}_0 = \alpha/2$ gives the exact optimising value for a *symmetric unimodal distribution*. Using this starting value ensures that the case of a symmetric distribution is solved exactly, without any iterations of the algorithm; this is a useful simplification that ensures that the algorithm is simple and precise for symmetric distributions. We recommend using this as a starting value, unless there are good reasons to the contrary, such as particular starting knowledge allowing a better placement within the admissible range.

The above framing presents the HDR as a constrained optimisation problem over the interval $0 \leq \theta \leq \alpha$, but it is possible to remove the constraint on the parameter range using a simple transformation method. To do this we take a twice-differentiable strictly increasing function $\phi \mapsto \theta$ over the unconstrained domain $\phi \in \mathbb{R}$ and with limiting values:

$$\theta(-\infty) = 0 \qquad \theta(\infty) = \alpha.$$

Our objective function for the width of the interval can then be written directly in terms of the unconstrained parameter $\phi$ with the following form:

$$W(\phi) = Q(\theta(\phi) + 1 - \alpha) - Q(\theta(\phi)),$$
$$W'(\phi) = \theta'(\phi)W'(\theta(\phi)),$$
$$W''(\phi) = \theta''(\phi)W'(\theta(\phi)) + \theta'(\phi)^2 W''(\theta(\phi)).$$

There are many possible choices for the unconstrained parameter, but one useful form is the smooth transformation $\phi = \log(\theta) - \log(\alpha - \theta)$. This transformed unconstrained parameter gives the following forms for the lower-probability function and its derivatives:

$$\theta(\phi) = \frac{\alpha}{1 + \exp(-\phi)},$$

$$\theta'(\phi) = \frac{\theta(\phi)}{1 + \exp(\phi)},$$

$$\theta''(\phi) = \theta'(\phi) \cdot \frac{1 - \exp(2\phi)}{1 + 2\exp(\phi) + \exp(2\phi)}.$$

Taking $\hat{\phi}_0 = 0$ as the starting value of the unconstrained parameter in the iterative algorithm gives the recommended value of $\hat{\theta}_0 = \alpha/2$ for the corresponding lower-tail probability.

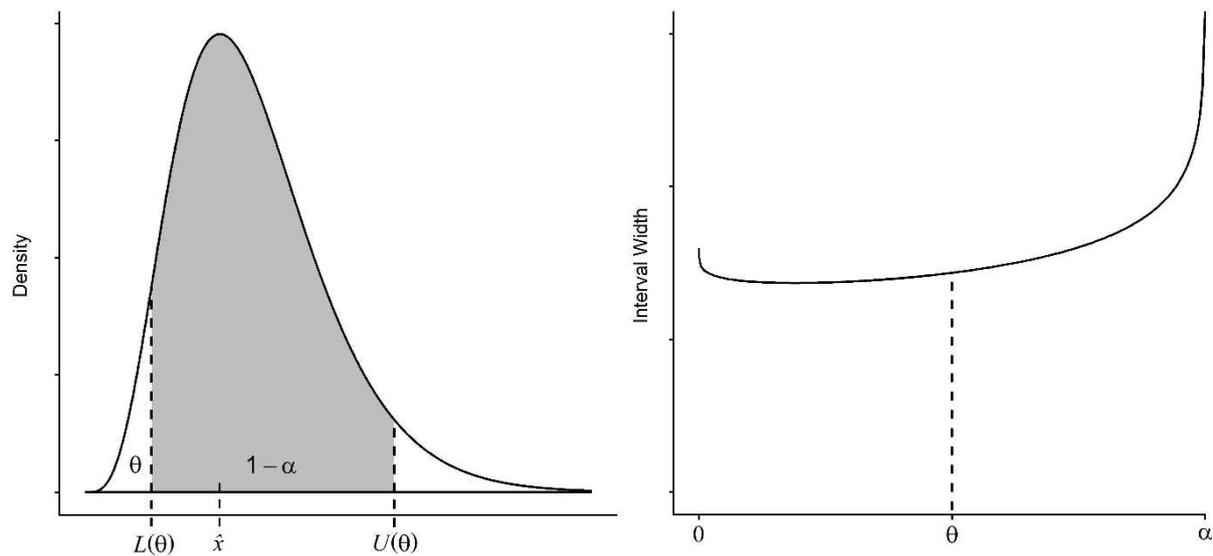

**FIGURE 1:** Density plot and corresponding width function (this interval is not the HDR)

In Figure 1 above we show the HDR optimisation problem for a strictly quasi-concave density. The grey area in the plot is the required coverage probability $1 - \alpha$ and the area to the left of this is the parameter $\theta$. In this particular case we have $\alpha \leq F(\hat{x}) \leq 1 - \alpha$ so the all values of $\theta$ in the range give bounds that span the mode (i.e., the admissible range is the full range). This means that the width function is strictly concave over the entire parameter range. As can be seen, the present interval is not the HDR, since it does not minimise the width function. The HDR is obtained by moving the parameter $\theta$ to the left until it minimises the width function, at which point the density will be equal at the lower and upper bounds.

## 5. Coding the HDR algorithm in `R` — monotone or quasi-concave densities

We will code the above algorithms to compute the HDR in the case of a univariate continuous unimodal distribution (i.e., where the density is either monotone or strictly quasi-concave). The monotone case includes a some standard continuous distributions, such as the exponential distribution, and certain cases of the gamma, Weibull, and Snedecor F distributions. The non-monotonic strictly quasi-concave case includes several other standard continuous distributions, such as the normal, Student's T, chi-squared, log-normal, and most parameterisations of the gamma, Weibull, beta, and Snedecor F distributions.

To compute these HDRs, we created the functions `HDR.monotone` and `HDR.unimodal` shown in the code below.[12] These functions take inputs for the significance level $\alpha$ and the quantile function of the distribution for which the HDR is to be computed; the user also has the option to input the density function and the logarithmic-derivative-density function.[13] The monotone case is a simple function, which computes the HDR from the quantile function, plus a specification of the direction of monotonicity. The unimodal function computes the HDR using the iterative nonlinear optimisation algorithm in the `nlm` function (see Schnabel, Koontz and Weiss 1985). As discussed above, we set the starting parameter for this algorithm to be the exact optima for a symmetric distribution, which means that the algorithm gives an exact HDR in this case, without having to generate iterations of the optimising algorithm. In addition to the required probability functions, our function also allows the user to specify a textual description of the distribution, and specify parameters for the nonlinear optimisation algorithm, including the gradient tolerance, step tolerance, and maximum number of iterations.

In all our functions, the HDR we compute is an interval using the standard structure in the `sets` package in `R` (Meyer and Hornik 2009). The output of both the `HDR.monotone` and

---

[12] For brevity, the functions we give in this paper do not include any checks on the inputs, to ensure that they are of the appropriate type, and to ensure that numerical values are in the admissible range. In actual implementation of these functions outside this paper we have added additional checks on inputs, but we omit those here for brevity. Note also that for consistency of the inputs between the two functions, we allow the input of the density function and the logarithmic-derivative-density function into `HDR.monotone`, even though these inputs are not used in that function.

[13] The quantile function of the distribution is a necessary input, but the density function and the logarithmic-derivative-density function may be omitted. The latter objects are used to generate analytic outputs for the first and second derivatives of the objective function. If these are omitted then these derivatives are approximated by numerical differentiation instead of being computed analytically from the formulae derived here. Tests by the author show that the function performs well in either case.

HDR.unimodal functions is an object with classes 'hdr' and 'interval' showing the HDR as an interval, with some additional attributes containing the coverage probability, a textual description of the distribution used, and information on the optimisation method (the number of iterations and the code for why the optimising algorithm terminated). The function allows the user to enter significance values $\alpha = 0$ or $\alpha = 1$. In the former case the HDR is the entire support of the distribution and in the latter case the HDR is the empty set. We accompany the HDR algorithm with a custom print function print.hdr to give user-friendly printing for hdr objects (see Appendix).

---

**Algorithm 1: Compute HDR for monotone density**

```
This function produces the HDR for a continuous distribution with a monotone
density function, at a specified level of significance.  The algorithm requires
the sets package.

Inputs:        The coverage probability cover.prob
               The quantile function Q
               A logical value indicating whether the density is decreasing
               The name of the distribution, called distribution
Output:        A 'hdr' object giving the HDR for the random variable.
```

```r
HDR.monotone <- function(cover.prob, Q, decreasing = TRUE,
                         distribution = 'an unspecified input distribution') {

  #Compute the HDR in trivial cases where cover.prob is 0 or 1
  #When cover.prob = 0 the HDR is the empty region
  if (cover.prob == 0) {
    HDR <- sets::interval()
    attr(HDR, 'method') <- as.character(NA) }
  #When cover.prob = 1 the HDR is the support of the distribution
  if (cover.prob == 1) {
    HDR <- sets::interval(l = Q(0), r = Q(1), bounds = 'closed')
    attr(HDR, 'method') <- as.character(NA) }

  #############

  #Compute the HDR in non-trivial cases where 0 < cover.prob < 1

  if ((cover.prob > 0) & (cover.prob < 1)) {
    if (decreasing) { L <- Q(0);          U  <- Q(cover.prob) } else {
                      L <- Q(1-cover.prob); U  <- Q(1) }
    HDR <- sets::interval(l = L, r = U, bounds = 'closed')
    attr(HDR, 'method') <- paste0('Computed using monotone optimisation') }

  #Add class and attributes
  class(HDR) <- c('hdr', 'interval')
  attr(HDR, 'probability')  <- cover.prob
  attr(HDR, 'distribution') <- distribution

  HDR }
```

---

**Algorithm 2: Compute HDR for unimodal density**

---

This function produces the HDR for a continuous distribution with a quasi-concave density function, at a specified level of significance. The algorithm requires the **sets** package.

**Inputs:**      The coverage probability **cover.prob**
             The quantile function **Q**
             The density function **f**
             The logarithmic-derivative density **u**
             The name of the distribution, called **distribution**
             Inputs **gradtol**, **steptol** and **iterlim** for the nlm optimisation
**Output:**     A 'hdr' object giving the HDR for the random variable.

---

```
HDR.unimodal <- function(cover.prob, Q, f = NULL, u = NULL,
                         distribution = 'an unspecified input distribution',
                         gradtol = 1e-10, steptol = 1e-10, iterlim = 100) {

  #Compute the HDR in trivial cases where cover.prob is 0 or 1
  #When cover.prob = 0 the HDR is the empty region
  if (cover.prob == 0) {
    HDR       <- sets::interval()
    attr(HDR, 'method') <- as.character(NA) }
  #When cover.prob = 1 the HDR is the support of the distribution
  if (cover.prob == 1) {
    HDR       <- sets::interval(l = Q(0), r = Q(1), bounds = 'closed')
    attr(HDR, 'method') <- as.character(NA) }

  #Compute the HDR in non-trivial cases where 0 < cover.prob < 1
  #Computation is done using nonlinear optimisation using nlm

  if ((cover.prob > 0) & (cover.prob < 1)) {

  #Set objective function
  WW <- function(phi) {

    #Set parameter functions
    T0 <- (1-cover.prob)/(1+exp(-phi))
    T1 <- T0/(1+exp(phi))
    T2 <- T1*((1-exp(2*phi))/(1+2*exp(phi)+exp(2*phi)))

    #Set interval bounds and objective
    L  <- Q(T0)
    U  <- Q(T0 + cover.prob)
    W0 <- U - L

    #Set gradient and Hessian of objective (if able)
    if (!is.null(f)) {
      attr(W0, 'gradient') <- T1*(1/f(U) - 1/f(L)) }
    if (!is.null(f) & !is.null(u)) {
      attr(W0, 'hessian')  <- T2*(1/f(U) - 1/f(L)) +
                              T1^2*(u(L)/(f(L)^2) - u(U)/(f(U)^2)); }

    W0 }

  #Compute the HDR
  #The starting value for the parameter phi is set to zero
  #This is the exact optima in the case of a symmetric distribution
  OPT <- nlm(WW, p = 0,
             gradtol = gradtol, steptol = steptol, iterlim = iterlim)
  TT <- (1-cover.prob)/(1+exp(-OPT$estimate))
  L   <- Q(TT)
  U   <- Q(TT + cover.prob)
  HDR <- sets::interval(l = L, r = U, bounds = 'closed')
```

```
    #Description of method
    METHOD <- ifelse((OPT$iterations == 1),
              paste0('Computed using nlm optimisation with ',
                    OPT$iterations, ' iteration (code = ', OPT$code, ')'),
              paste0('Computed using nlm optimisation with ',
                    OPT$iterations, ' iterations (code = ', OPT$code, ')'))

    #Add class and attributes
    class(HDR) <- c('hdr', 'interval')
    attr(HDR, 'method') <- METHOD; }
    attr(HDR, 'probability')  <- cover.prob
    attr(HDR, 'distribution') <- distribution

    HDR }
```

The base functions `HDR.monotone` and `HDR.unimodal` take fully specified probability functions that have fixed parameter values, so that those functions operate as functions with a single scalar argument. We can use these functions as a basis for creating HDR functions that are customised to particular families of univariate continuous unimodal distributions, allowing the specification of parameters and customised textual information about the distribution. In many of the base distributions in **R** there are some parameterisations that give a monotone density function and other parameterisations that give a non-monotone unimodal density function. For a distributional family of this kind, we can therefore create a "wrapper" function that takes in a significance level and parameter values, and uses these to call on the underlying functions given above to produce a HDR using an appropriate method that depends on the parameters of the distribution.

Below we illustrate this method by creating a function `HDR.chisq` that computes HDRs for the family of chi-squared distributions, where the parameters of the distribution (degrees-of-freedom and non-centrality parameter) are inputs to the function. This "wrapper" function sets scalar quantile and density functions, sets the textual description of the distribution, and then computes the resulting HDR by calling the underlying functions for monotone and unimodal cases. The function computes the HDR from a monotone decreasing optimisation in the case where the distribution is monotone (i.e., degrees-of-freedom no greater than two) or from the nonlinear optimisation in the case where the distribution is non-monotone but unimodal (i.e., degrees-of-freedom greater than two). The output of the `HDR.chisq` function is an interval with the classes and information we specified above for our main function. The name of the distribution is given textually, and the `'hdr'` object contains the coverage probability, distribution, and information on the convergence of the optimisation algorithm.

> **Algorithm 3: Compute HDR for chi-squared distribution**
>
> This function produces the HDR for a chi-squared distribution, at a specified
> level of significance.  The algorithm requires the **sets** package.
>
> **Inputs:**      The coverage probability **cover.prob**
>                  The degrees-of-freedom parameter **df**
>                  The non-centrality parameter **ncp**
>                  Inputs **gradtol**, **steptol** and **iterlim** for the nlm optimisation
> **Output:**      A 'hdr' object giving the HDR for the chi-squared distribution.

```r
HDR.chisq <- function(cover.prob, df, ncp = 0,
                      gradtol = 1e-10, steptol = 1e-10, iterlim = 100) {

  #Generate probability functions with stipulated parameters
  QQ <- function(L) { qchisq(L, df, ncp) }
  DD <- function(L) { dchisq(L, df, ncp) }

  #Set text for distribution
  DIST <- ifelse(ncp == 0,
           paste0('chi-squared distribution with ', df, ' degrees-of-freedom'),
           paste0('chi-squared distribution with ', df,
                  ' degrees-of-freedom and non-centrality parameter = ', ncp))

  #Compute HDR in monotone case
  if (df <= 2) {
    HDR <- HDR.monotone(cover.prob, Q = QQ, f = DD, distribution = DIST,
                        decreasing = TRUE) }

  #Compute HDR in unimodal case;
  if (df > 2) {
    HDR <- HDR.unimodal(cover.prob, Q = QQ, f = DD, distribution = DIST,
                        gradtol = gradtol, steptol = steptol,
                        iterlim = iterlim) }

HDR }
```

An example of the use of this function is shown below.  The printed output of the function is a user-friendly statement detailing the optimisation method, and the interval that was generated. This example shows a HDR with 98% coverage probability for the chi-squared distribution with 30 degrees-of-freedom.  Our subsequent computations show the density values at the bounds of the HDR, and confirm that these are equal to each other, subject to mild inaccuracy (this accuracy can be altered by changing the `gradtol` and `steptol` parameters in the HDR function, and this affects the nonlinear optimisation algorithm).

```r
#Generate and print a HDR
(HDR <- HDR.chisq(cover.prob = 0.98, df = 30, ncp = 0))

        Highest Density Region (HDR)

98.00% HDR for chi-squared distribution with 30 degrees-of-freedom
Computed using nlm optimisation with 6 iterations (code = 1)

[13.9324865197342, 49.3372669844555]
```

```
#Show the density at the bounds
DENS_LOWER <- dchisq(min(HDR), df = 30, ncp = 0)
DENS_UPPER <- dchisq(max(HDR), df = 30, ncp = 0)
c(DENS_LOWER, DENS_UPPER)

[1] 0.003428795 0.003428795

#Show the difference in density values
DENS_UPPER - DENS_LOWER

[1] -1.708703e-16
```

It is simple to program analogous functions for other standard distributions in base **R** or its extension packages, so long as these families of distributions have densities that are either monotone or quasi-concave over their parameter values. Testing by the author shows that the `HDR.unimodal` function operates well on various standard unimodal input distributions in the `stats` package in **R**. Over a range of tests the function finds the HDR to a high degree of accuracy with rapid computation.[14]

**6. HDRs for quasi-convex densities**

In this section we will consider the case of a continuous density that is strictly quasi-convex over its support, so that it is bimodal. This third case occurs for some standard distributional forms over certain parameter values (e.g., using a beta distribution with both scale parameters below one). To frame this case we will consider a distribution with support $[L_*, U_*]$ so that the values $L_* = Q(0)$ and $U_* = Q(1)$ are the local maximums of the bimodal density. In this case, the HDR will be a closed interval of the form $[L_*, L] \cup [U, U_*]$. An interval with fixed coverage probability $1 - \alpha$ must have lower probability $0 \leq F(L) \leq 1 - \alpha$ and so the region can be fully characterised by the parameter $\theta \equiv F(L)$ in the range $0 \leq \theta \leq 1 - \alpha$. The values $L$ and $U$ can now be written in terms of this parameter as:

$$L(\theta) = Q(\theta) \qquad U(\theta) = Q(\theta + \alpha).$$

The resulting "width" of the region (which is now actually the sum of the widths of the lower and upper parts of the region) can be written in terms of the parameter as:

$$W(\theta) = 1 - Q(\theta + \alpha) + Q(\theta) \qquad \text{for all } 0 \leq \theta \leq 1 - \alpha.$$

---

[14] For example, the author was able to use a looped calculation to compute the HDR intervals for the chi-squared distribution over 10,000 different values of the degrees-of-freedom parameter. This computation took 5.76 second on a standard personal computer (approximately 1,735 HDR outputs per second).

As in our previous optimisation, since the coverage probability is fixed, we can find the HDR by minimising this objective function over the constrained range of the parameter $\theta$. We then substitute the optimising parameter to obtain the lower and upper bounds of the HDR. The objective function has first and second derivatives:

$$W'(\theta) = \frac{1}{f(L(\theta))} - \frac{1}{f(U(\theta))},$$

$$W''(\theta) = \frac{u(U(\theta))}{f(U(\theta))^2} - \frac{u(L(\theta))}{f(L(\theta))^2}.$$

Since the density function $f$ is assumed to be strictly quasi-convex, the density has a unique minimising point $\hat{x}$ in its support and its slope satisfies the equation $\operatorname{sgn} u(x) = \operatorname{sgn}(x - \hat{x})$ for all $x \in \mathbb{R}$. Thus, for any parameter value $\theta$ giving bounds $L(\theta) < \hat{x} < U(\theta)$ that span the minimising value, we have $u(L(\theta)) < 0$ and $u(U(\theta)) > 0$ which implies that $W''(\theta) > 0$. We again refer to this range of parameter values as the "admissible range", noting that in this case the relevant value $\hat{x}$ is the minimising value rather than the mode. As with the previous optimisation problem, we see that the width function is strictly convex over the admissible range of parameter values $\theta$.

**THE "ADMISSIBLE RANGE" FOR STRICTLY QUASI-CONVEX DENSITY:** When dealing with a strictly quasi-convex function that is not monotonic, the bounds of the HDR must always span the minimising point in the support, which gives the inequality $L(\theta) < \hat{x} < U(\theta)$ defining the admissible range. For any value $0 < \alpha < 1$ the admissible range of $\theta$ is the range:

$$\max(0, F(\hat{x}) - \alpha) < \theta < \min(1 - \alpha, F(\hat{x})).$$

(This result is obtained by inverting the pair of inequalities defining the admissible range, and overlaying the result onto the additional requirement that $0 \leq \theta \leq \alpha$.) In the sections outside the admissible range the width function is monotonic, but it is not necessarily convex. It is also notable that if $1 - \alpha \leq F(\hat{x}) \leq \alpha$ then the admissible range is the full range $0 \leq \theta \leq 1 - \alpha$. So long as the mode is not too close to one end of the support, and so long as $\alpha$ is not too small, this latter inequality will be satisfied. In most HDR calculations of interest in practice, the admissible range is the full range. □

The above results show that the width function is strictly concave over the "admissible range" of parameter values that lead to bounds spanning the minimising point of the density. (To avoid any confusion, note that the density function is still strictly *quasi-convex*, but the "width" function is strictly *convex* over the admissible range.) The optimisation problem therefore has unique critical point $\hat{\theta}$ that is the minimising parameter, and this gives values $\hat{L} < \hat{x} < \hat{U}$ that span the minimising point $\hat{x}$ in the support of the distribution. The bounds are characterised by the first-order condition $f_* = f(\hat{L}) = f(\hat{U})$ just as in the quasi-concave case, which is the well-known intuitive condition for a HDR. Since the width function is strictly convex over the admissible range, the second order condition for a local minimum is also satisfied.

All subsequent optimisation issues are analogous to the quasi-concave case dealt with in the previous section. As with the quasi-concave case, solution to the critical point equation will generally iterative methods. The optimisation can be converted to an unconstrained problem using an appropriate transform $\phi \mapsto \theta$, as described in the previous section. (We now use $\theta(\phi) = (1-\alpha)/1 + \exp(-\phi)$ but all other equations are the same.) We again recommend using the starting value $\hat{\theta}_0 = \alpha/2$, since this is the optima for a symmetric distribution.

As can be seen from the above working, although the objective function in the quasi-convex case is different to the quasi-concave case, but both cases lead to a well-behaved optimisation problem over the respective parameter ranges, where the objective function is strictly convex over the "admissible range" yielding values $L(\theta) < \hat{x} < U(\theta)$. In the quasi-concave case the value $\hat{x}$ is the mode of the density function and in the quasi-convex case it is the minimising value of the density function over its support. These differences do not affect the optimisation problem in terms of its iterative convergence to the optima.

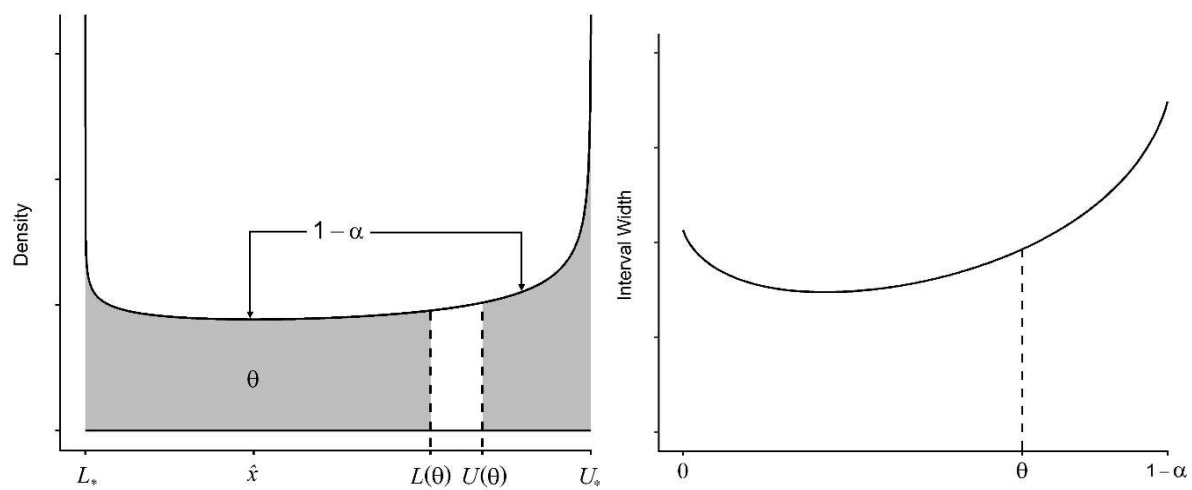

**FIGURE 2:** Density plot and corresponding width function (this interval is not the HDR)

In Figure 2 above we show the HDR optimisation problem for a strictly quasi-convex density. The grey area is the required coverage probability $1 - \alpha$ and the area to the left is the parameter $\theta$. In this particular case the interval is outside the admissible range (both interval bounds are to the right of the minimising point in the support) but the width function is strictly convex nonetheless. As can be seen, the present interval is not the HDR, since it does not minimise the width. The HDR is obtained by moving the parameter $\theta$ to the left until it minimises the width function, at which point the density will be equal at the lower and upper bounds.

## 7. Coding the HDR algorithm in R —quasi-convex densities

We will code the above algorithm to compute the HDR in the case of a univariate continuous bimodal distribution that has a strictly quasi-convex density function. This case captures the beta distribution over a small subset of its parameter range. To compute this HDR, we create the function `HDR.bimodal` shown in the code below. As before, this function takes inputs for the significance level $\alpha$ and the quantile function of the distribution for which the HDR is to be computed; the user also has the option to input the density function and the logarithmic-derivative-density function, and the function computes the HDR using the iterative nonlinear optimisation algorithm in the `nlm` function. As discussed above, we set the starting parameter for this algorithm to be the exact optima for a symmetric bimodal distribution, which means that the algorithm gives an exact HDR in this case, without having to generate iterations of the optimising algorithm. The function takes the same inputs as in the unimodal case, including specification of parameters for the `nlm` optimisation.

```
                Algorithm 4: Compute HDR for bimodal density

This function produces the HDR for a continuous distribution with a quasi-convex
density function, at a specified level of significance.  The algorithm requires
the sets package.

Inputs:      The coverage probability cover.prob
             The quantile function Q
             The density function f
             The logarithmic-derivative density u
             The name of the distribution, called distribution
             Inputs gradtol, steptol and iterlim for the nlm optimisation
Output:      A 'hdr' object giving the HDR for the random variable.

HDR.bimodal <- function(cover.prob, Q, f = NULL, u = NULL,
                        distribution = 'an unspecified input distribution',
                        gradtol = 1e-10, steptol = 1e-10, iterlim = 100) {
```

```r
  #Compute the HDR in trivial cases where cover.prob is 0 or 1
   #When cover.prob = 0 the HDR is the empty region
  if (cover.prob == 0) {
    HDR <- sets::interval()
    attr(HDR, 'method') <- as.character(NA) }
 #When cover.prob = 1 the HDR is the support of the distribution
  if (cover.prob == 1) {
    HDR <- sets::interval(l = Q(0), r = Q(1), bounds = 'closed')
    attr(HDR, 'method') <- as.character(NA) }

  #Compute the HDR in non-trivial cases where 0 < cover.prob < 1
  #Computation is done using nonlinear optimisation using nlm

  if ((cover.prob > 0) & (cover.prob < 1)) {

  #Set objective function
  WW <- function(phi) {

    #Set parameter functions
    T0 <- cover.prob/(1+exp(-phi))
    T1 <- T0/(1+exp(phi))
    T2 <- T1*((1-exp(2*phi))/(1+2*exp(phi)+exp(2*phi)))

    #Set interval bounds and objective
    L  <- Q(T0)
    U  <- Q(T0 + 1 - cover.prob)
    W0 <- 1 - U + L

    #Set gradient of objective (if able)
    if (!is.null(f)) {
      attr(W0, 'gradient') <- - T1*(1/f(U) - 1/f(L)) }

    #Set Hessian of objective (if able)
    if (!is.null(f) & !is.null(u)) {
      attr(W0, 'hessian')  <- - T2*(1/f(U) - 1/f(L)) -
                  T1^2*(u(L)/(f(L)^2) - u(U)/(f(U)^2)) }

    W0 }

  #Compute the HDR
  #The starting value for the parameter phi is set to zero
  #This is the exact optima in the case of a symmetric distribution
  OPT <- nlm(f = WW, p = 0,
             gradtol = gradtol, steptol = steptol, iterlim = iterlim);
  TT <- (1-cover.prob)/(1+exp(-OPT$estimate))
  L  <- Q(TT*cover.prob/(1-cover.prob))
  U  <- Q(TT*cover.prob/(1-cover.prob) + 1 - cover.prob)
  HDR1 <- sets::interval(l = Q(0), r = L, bounds = 'closed')
  HDR2 <- sets::interval(l = U, r = Q(1), bounds = 'closed')
  HDR  <- sets::interval_union(HDR1, HDR2)

  #Add the description of the method
  METHOD <- ifelse((OPT$iterations == 1),
            paste0('Computed using nlm optimisation with ',
                    OPT$iterations, ' iteration (code = ', OPT$code, ')'),
            paste0('Computed using nlm optimisation with ',
                    OPT$iterations, ' iterations (code = ', OPT$code, ')'))
  attr(HDR, 'method') <- METHOD }

  #Add class and attributes
  class(HDR) <- c('hdr', 'interval')
  attr(HDR, 'probability')  <- cover.prob
  attr(HDR, 'distribution') <- distribution

  HDR }
```

Using the function `HDR.bimodal`, we can now program HDR functions for any family of distributions that has densities that are either monotone, quasi-concave, or quasi-convex over their parameter values. As an example, we create a function `HDR.beta` that generates HDRs for the beta distribution, which encompasses all three cases; the monotonic density, quasi-concave density, and quasi-convex density.

```
                 Algorithm 5: Compute HDR for beta distribution
```
```
This function produces the HDR for a beta distribution, at a specified level of
significance.  The algorithm requires the sets package.

Inputs:       The coverage probability cover.prob
              The shape parameter shape1 and shape2
              The non-centrality parameter ncp
              Inputs gradtol, steptol and iterlim for the nlm optimisation
Output:       A 'hdr' object giving the HDR for the chi-squared distribution.
```

```
HDR.beta <- function(cover.prob, shape1, shape2, ncp = 0,
                     gradtol = 1e-10, steptol = 1e-10, iterlim = 100) {

  #Simplify probability functions (with stipulated parameters)
  QQ <- function(L) { qbeta(L, shape1, shape2, ncp) }
  DD <- function(L) { dbeta(L, shape1, shape2, ncp) }

  #Set text for distribution
  DIST <- ifelse(ncp == 0,
          ifelse(((shape1 == 1) && (shape2 == 1)),
                 'standard uniform distribution',
            paste0('beta distribution with shape1 = ', shape1,
                   ' and shape2 = ', shape2)),
            paste0('beta distribution with shape1 = ', shape1,
                   ' and shape2 = ', shape2,
                   ' and non-centrality parameter = ', ncp))

  #Compute HDR in monotone cases
  if ((shape1 <= 1) && (shape2 >  1)) {
    HDR <- HDR.monotone(cover.prob, Q = QQ, f = DD, distribution = DIST,
                        decreasing = TRUE) }
  if ((shape1 >  1) && (shape2 <= 1)) {
    HDR <- HDR.monotone(cover.prob, Q = QQ, f = DD, distribution = DIST,
                        decreasing = FALSE) }

  #Compute HDR in uniform case
  if ((shape1 == 1) && (shape2 == 1)) {
    HDR <- HDR.unimodal(cover.prob, Q = QQ, f = DD, distribution = DIST,
                        gradtol = gradtol, steptol = steptol, iterlim = iterlim) }

  #Compute HDR in unimodal case
  if ((shape1 > 1) && (shape2 > 1)) {
    HDR <- HDR.unimodal(cover.prob, Q = QQ, f = DD, distribution = DIST,
                        gradtol = gradtol, steptol = steptol, iterlim = iterlim) }

  #Compute HDR in bimodal case
  if ((shape1 < 1) && (shape2 < 1)) {
    HDR <- HDR.bimodal(cover.prob, Q = QQ, f = DD, distribution = DIST,
                       gradtol = gradtol, steptol = steptol, iterlim = iterlim) }

  HDR }
```

It turns out that all of the families of continuous distributions in base R have density functions that are either monotonic, quasi-concave, or quasi-convex, over their support. The above functions are sufficient to create individualised HDR functions for each standard continuous distributional family. This is done in a manner analogous to Algorithms 3 and 5 above (which give functions for the chi-squared distribution and the beta distribution). To program an individualised HDR function for a distributional family, we take an input parameterisation for the distribution and then use the appropriate HDR algorithm for the shape of that distribution. Corresponding HDRs for discrete distributions are discussed in a later section.

**8. HDRs for other continuous densities**

The problem of computing HDRs for other densities —neither monotone, quasi-concave, or quasi-convex, over their support— can be handled using the "density quantile approach" by framing an appropriate optimisation problem using a number of intervals whose union is the HDR. This optimisation problem can be framed in a number of ways, but we will examine one framing that is a particularly natural extension of the above methods. In our method, we will break the density up into a piecewise function with parts that are monotone, quasi-concave, or quasi-convex. Once the density is viewed in piecewise form in this manner, the HDR for the whole density can be regarded as a union of HDRs for the conditional densities concentrated on each of the range segments. The HDR for the overall density can therefore be computed by solving a constrained optimisation problem where we specify coverage probabilities on each of the range segments, compute the HDRs on these segments, and then optimise the coverage probabilities subject to the requirement that they must add up to the total coverage probability. To see this technique a bit more clearly, we examine the most common case where we have a differentiable density function with a countable number of local minima, and no flat sections on its support.[15] Denote the local minima of the density by $x_1 < \cdots < x_m$ and consider the corresponding range segments $[x_k, x_{k+1}]$ for all $k = 0, \ldots, m$.[16] We will find the HDR for the overall density with coverage probability $1 - \alpha$. This can be written as the union:

---

[15] The same basic idea can be extended if the number of local minima is countably infinite, or if there are flat sections, etc. For non-continuous densities it may be more useful to use some sections that are quasi-convex. We will not give a full treatment here, but will instead show one broad case, to illustrate the general method.
[16] We set $x_0 = Q(0)$ and $x_{m+1} = Q(1)$ with corresponding changes to the openness of the interval if either of these values is infinite. It is also worth noting that there is overlap between these range segments at the boundary points. Since the HDR is defined as a closed interval, this does not cause problems, and indeed, it is the simplest way to frame the optimisation.

$$\mathcal{H} = \bigcup_{k=0}^{m} \mathcal{H}_k \qquad \mathcal{H}_k \subseteq [x_k, x_{k+1}].$$

To compute each HDR segments we define the values $\pi_0, \ldots, \pi_m$ and the conditional densities $f_0, \ldots, f_m$ for the range segments as:

$$\pi_k \equiv \mathbb{P}(x_k \leq X \leq x_{k+1}) \qquad f_k(x) \equiv \frac{f(x)}{\pi_k} \cdot \mathbb{I}(x_k \leq x \leq x_{k+1}).$$

For each range segment $k = 0, \ldots, m$ the value $\pi_k$ is the probabilities that the random variable falls in that range segment and the conditional density $f_k$ is the conditional density given that the random variable falls into that range segment. To find the HDR segments $\mathcal{H}_0, \ldots, \mathcal{H}_m$ we consider parameter vector $\boldsymbol{\alpha} = (\alpha_0, \ldots, \alpha_m) \in [0,1]^{m+1}$. For each range segment $k = 0, \ldots, m$ we find the HDR with coverage probability $1 - \alpha_k$ on the conditional density $f_k$. (Since each of these conditional densities is quasi-concave, we can use the standard algorithm for finding the HDR on a strictly quasi-concave density.) We define the resulting optimised width function as $\widehat{W}_k(\alpha_k) = \min_{0 \leq \theta_k \leq \alpha_k} W_k(\theta_k | \alpha_k)$ and we solve the constrained optimisation problem:

$$\text{Minimise} \qquad W(\boldsymbol{\alpha}) = \sum_{k=1}^{m} \widehat{W}_k(\psi_k) \qquad \text{subject to} \qquad \sum_{k=1}^{m} (1 - \alpha_k) \cdot \pi_k = 1 - \alpha.$$

Solving this constrained optimisation yields the optimum vector $\widehat{\boldsymbol{\alpha}}$ which then gives the HDR segments $\mathcal{H}_k = \left[ L(\widehat{\theta}_k | \widehat{\alpha}_k), U(\widehat{\theta}_k | \widehat{\alpha}_k) \right]$ which gives the resulting HDR:

$$\mathcal{H} = \bigcup_{k=0}^{m} \mathcal{H}_k = \bigcup_{k=0}^{m} \left[ L(\widehat{\theta}_k | \widehat{\alpha}_k), U(\widehat{\theta}_k | \widehat{\alpha}_k) \right].$$

This general optimisation problem has three parts. The first part is to find the local minimums of the function and convert the density to a piecewise form involving strictly quasi-concave parts on the range segments between the minimising points. The second part is the optimisation of HDRs on range segments where the conditional density is a strictly quasi-concave function, leading to a strictly convex width function over the admissible range; this is done using the method shown in the section on quasi-concave densities above. The third part is the constrained optimisation problem to find the coverage probabilities on each of the range segments, that add up to the total coverage probability. This latter optimisation problem uses the optimised width functions for the HDRs on the range segments as an input. These steps are shown in Figure 3.

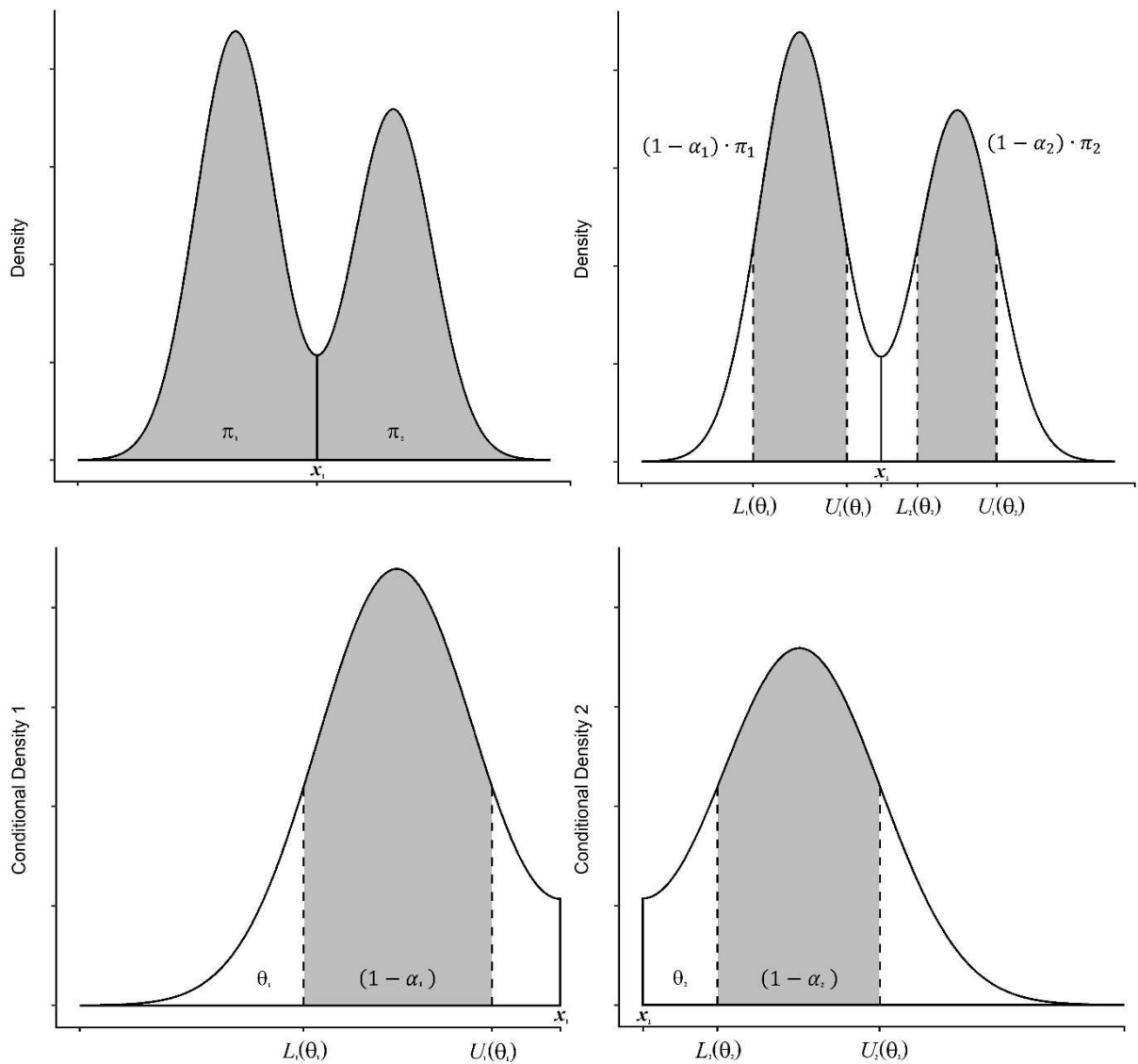

**FIGURE 3:** Density plot for bimodal density, split into piecewise conditional densities

In Figure 3 we show the parts of the general optimisation process for a bimodal density function what is neither strictly quasi-concave or quasi-convex. Starting with the figure in the top left, we determine the local minimum $x_1$ and determine the probabilities $\pi_1$ and $\pi_2$ for the two range segments. In the two figures below we take each of the conditional densities (which are strictly quasi-concave) and form HDRs on those densities for a stipulated coverage probability. We program this into a function of the coverage probability. Finally, in figure in the top right, we optimise the coverage probabilities for the two range segments, under the constraint that they add up to the total coverage probability. (This leads to equal densities at all the bounds.)

It is worth noting here that in cases where the density cut-off for the HDR is equal to or below the density of a local minimising point, then the HDR will encompass that minimising point, and so the boundaries of the intervals on the two range segments on either side will be at the minimising point. Suppose we consider the minimising point $x_k$ that separates range segments $k$ and $k + 1$. If $f(x_k) \geq f_*$ (i.e., if the density at the minimising point is above the density cut-off of the HDR) then we can collapse range segments $k$ and $k + 1$ into a single segment of the form $[L_k, U_{k+1}]$ (which encompasses the point $U_k = x_k = L_{k+1}$). For this reason, it may be fruitful to add a preliminary step to the algorithm, where we compute the HDRs at each of the cut-off points $f_* = f(x_1), \ldots, f(x_m)$ in order to determine which of the minimising points is encompassed by the HDR at the specified required coverage probability. By performing this additional preliminary step, we can reduce the number of range segments and thereby obtain a simpler algorithm for the remainder of the HDR algorithm. We also ensure that the minimising points with densities above the cut-off density are not excluded from the HDR due to issues with the numerical computation in the optimisation.[17]

The standard univariate distributions in **R** do not include a distribution of this form, but we have included the method for completeness over all univariate distributions. Programming the method requires multiple steps, including an initial step to compute local minima of the density, a subsequent optimisation step to compute the HDRs across each conditional density on each range segment, and a final step to combine these into a single HDR. Programming this analysis is beyond the scope of the present paper.

## 9. HDRs for discrete distributions

The above analysis is for continuous density functions and so we use calculus methods to solve the relevant optimisation problem for the HDR. In the case of discrete distributions, we obtain a discrete optimisation problem that is solved using analogous methods in discrete calculus. This can be done for monotone densities, unimodal (quasi-concave) densities, bimodal (quasi-convex) densities, and so on. Care must be taken when translating the continuous algorithms into their discrete counterparts, since the endpoints of the intervals in the discrete case have a positive point-mass. Indeed, if one attempts to apply the methods for continuous densities to discrete distributions without any accounting for the probability mass of the end-points, this can lead to incorrect outputs that are not true HDRs. (We will show soon that this happens in some **R** packages that attempt to compute HDRs in highly general cases.)

---

[17] This step is not be necessary if the optimisation performed in the second step can go right up to the boundary of the minimising point. However, it is important to note that if the optimisation in this step is transformed to an unconstrained optimisation, then the computed optima will not go right up to the boundary, and so the HDR will appear to have tiny gaps around the minimising points, which should not be present. For this reason we do recommend adding the preliminary step to remove unnecessary segmentation of the support.

One major advantage of dealing with discrete distributions is that it is possible to program an algorithm that computes the HDR accurately in the general case, without requiring knowledge of the "shape" of the probability mass function. This discrete optimisation problem and the resulting iterative method for solution are discussed in detail in O'Neill (2021). That paper develops an algorithm to accurately compute the HDR for any discrete distribution with support concentrated on a known countable set (e.g., the integers).[18] This is an iterative method that goes through the sequence of points on the countable set containing the support, and updates the HDR by considering one point at a time. The algorithm terminates when the unsearched points have total probability mass no larger than the lowest-probability point in the HDR.

We do not explore HDRs for discrete distributions in detail in this paper. Suffice to say, it is possible to use the iterative algorithm in O'Neill (2021) to find the HDR for any distribution with support on a known countable set, and it is possible to "discretise" the present continuous methods to give faster algorithms for discrete distributions with known shape —i.e., with a density function known to be monotone, unimodal, etc. In the **stat.extend** package these are implemented with a general function **HDR.discrete** what can be applied to any discrete distribution, and a more specific function **HDR.discrete.unimodal** that can be applied to any unimodal (including monotone) discrete distribution. These are applied to the discrete distributions in **R** to give HDR functions for those distributions. Combined with the above methods for continuous distributions, this allows accurate computation of HDRs for all the standard univariate distributions in **R**.

HDRs for mixtures of discrete and continuous distributions are also a simple extension to the results for continuous and discrete distributions separately, since discrete points always have preference over continuous intervals. Computation of these HDRs uses a simple comparison between the known mixture probabilities and the minimum coverage probability for the HDR; if the probability mass for the discrete part is no greater than the minimum coverage probability then the result is a HDR for the discrete part of the distribution; if the probability mass for the discrete part is greater than the minimum coverage probability then the entire set of discrete points is included in the HDR and the remaining continuous part is obtained using the present methods with an adjusted coverage probability.

---

[18] If the support does not fall on a known countable set then one cannot really get started with the problem because the search space is uncountable.

**10. Generalisation to method for finding optimal confidence intervals**

Our analysis of the optimisation problem in HDRs is also broadly applicable to other types of statistical problems involving the computation of interval estimates, in cases where one wishes to minimise the length of the resulting interval. In particular, the above optimisations can be employed to compute shortest-length confidence intervals in a range of cases where the interval can be expressed by a known density function. Confidence intervals for unknown parameters are widely used in applied statistics and have been recommended as a preferable alternative to reporting p-values of hypothesis tests (see e.g., Gardner and Altman 1986). Since confidence intervals are constructed to have a specified coverage probability under repeated sampling, and since all valid confidence intervals have this property, it is common to wish to select the interval that has the narrowest range, so as to make the most accurate inference possible at the specified level of confidence.

We will examine the general problem where we have a confidence interval constructed from a pivotal quantity using a monotone function of a variable of interest. We will consider a pivotal quantity for the parameter of interest $\psi$ which is also a function of an observable value $X$. The pivotal quantity we consider is $h_X(\psi) \sim$ Dist where $h$ is a strictly *decreasing* function and the resulting distribution has quantile function $Q$ that does not depend on the parameter of interest. This analysis can easily be extended to all monotonic functions.[19]

Our goal is to form a confidence interval for the parameter $\psi$ by first forming a probability interval for this parameter, with the required probability $1 - \alpha$. To form an interval with the required coverage probability we can again use the lower bound $L(\theta) = Q(\theta)$ and the upper bound $U(\theta) = Q(\theta + 1 - \alpha)$ where $\theta$ is a variable lower-tail probability that we can optimise later. Using these values we can form a probability interval for the pivotal quantity and the "invert" this to obtain an appropriate probability interval for the parameter of interest. Letting $g_X$ be the inverse function of $h_X$ (and noting that this is also strictly decreasing) we have:

$$1 - \alpha = \mathbb{P}(L(\theta) \leq h_X(\psi) \leq U(\theta))$$
$$= \mathbb{P}(g_X(U(\theta)) \leq \psi \leq g_X(L(\theta))).$$

---

[19] The case we examine is for a pivotal quantity based on a function $h_X$ and its inverse $g_X$ are strictly decreasing. The same analysis can be repeated for a strictly *increasing* function $h_X$ (so that $g_X$ is also strictly increasing). In this case, we obtain the probability interval $1 - \alpha = \mathbb{P}(g_X(L(\theta)) \leq \psi \leq g_X(U(\theta)))$ (i.e., the lower and upper bounds are switched), giving the width function $W(\theta) = g_X(U(\theta)) - g_X(L(\theta))$. Thus, the width function and its derivatives are the same as shown in the main analysis, except that they are all negated, and the necessary and sufficient condition for the critical point $\hat{\theta}$ to be a local minimum is the reverse of the inequality shown.

This gives us a probability interval with the required coverage, and substitution of the observed value of $X$ gives the corresponding confidence interval. The interval has width function:

$$W(\theta) = g_X(L(\theta)) - g_X(U(\theta)) \qquad \text{for all } 0 \leq \theta \leq \alpha.$$

This width function has first and second derivatives:

$$W'(\theta) = \frac{g'_X(L(\theta))}{f(L(\theta))} - \frac{g'_X(U(\theta))}{f(U(\theta))},$$

$$W''(\theta) = \frac{g''_X(L(\theta)) - u(L(\theta)) \cdot g'_X(L(\theta))}{f(L(\theta))^2} - \frac{g''_X(U(\theta)) - u(U(\theta)) \cdot g'_X(U(\theta))}{f(U(\theta))^2}.$$

This width function is a generalisation of the case we initially encountered when dealing with a unimodal density — that case occurs in the special case where $h_X$ (and therefore also $g_X$) is the identity function. The assumptions required to ensure strict convexity of the width function are stronger in the present case, so we will instead examine weaker assumptions allowing us to conclude that the width function is strictly quasi-convex, which is enough to ensure existence of a unique critical point optimising the confidence interval. To facilitate this analysis, define the function $m_X(\psi) \equiv g''_X(\psi)/g'_X(\psi)$. Since $g''_X(\psi) = m_X(\psi) \cdot g'_X(\psi)$ the second derivative of the width function can be written as:

$$W''(\theta) = \frac{m_X(L(\theta)) - u(L(\theta))}{f(L(\theta))} \cdot \frac{g'_X(L(\theta))}{f(L(\theta))} - \frac{m_X(U(\theta)) - u(U(\theta))}{f(U(\theta))} \cdot \frac{g'_X(U(\theta))}{f(U(\theta))}.$$

At any critical point $\hat{\theta}$ giving bounds $\hat{L}$ and $\hat{U}$ we have $g'_X(\hat{L})/f(\hat{L}) = g'_X(\hat{U})/f(\hat{U}) > 0$ so the second derivative at the critical point is given by:

$$W''(\hat{\theta}) = \text{pos. const} \times \left[ \frac{m_X(\hat{L}) - u(\hat{L})}{f(\hat{L})} - \frac{m_X(\hat{U}) - u(\hat{U})}{f(\hat{U})} \right].$$

Thus, a necessary and sufficient condition for the critical point $\hat{\theta}$ to be a local minimum is:

$$\frac{m_X(\hat{L}) - u(\hat{L})}{f(\hat{L})} > \frac{m_X(\hat{U}) - u(\hat{U})}{f(\hat{U})}.$$

This is a weaker assumption than is required to ensure strict convexity of the width function, but it is a stronger condition than what we required in our previous analysis of the unimodal HDR, where there was no intermediary transform. In some problems, this condition will hold over all values of $\theta$ in the analysis, or at least over all values in an "admissible range" that encompasses all possible critical points (i.e., not just at the critical points themselves). In this case, we know that all the critical points are local minimums, so we can apply a variation of the "only critical point in town test" to conclude that there is a unique critical point, which is the global minimising point (i.e., the optimising point that gives the shortest confidence interval for $\psi$). Computation of the optima proceeds as usual, by using numerical methods to find the unique critical point of the width function.

In some cases, the data in the pivotal quantity is "separable" from the main function, in the sense that $h_X(\psi) = k_X \cdot h(\psi)$ where $h$ is some fixed function and $k_X$ is a function of the data. In this case, letting $g$ be the inverse of $h$, we have $k_X \cdot W(\theta) = g(L(\theta)) - g(U(\theta))$ so that the width function is proportionate to a distance that does not depend on the data. In this case we can remove the constant $k_X$ from consideration and minimise the scaled width function $W(\theta)/k_X$ without consideration of the data. (We need to substitute $k_X$ back in at the end to get the actual bounds of the function, but it needn't be considered in the optimisation step.) All of the above functions with a subscript $X$ can then be rewritten without this subscript, since they do not depend on the data.

Taking any function of a pivotal quantity that does not depend on the parameter of interest or the data leads to another pivotal quantity. Consequently, optimal confidence interval problems can be framed in an infinite number of different ways, based on any pivotal quantity. By far the simplest case occurs when $h_X(\psi) = k_X \cdot \psi$ is a linear function, which means that the data is "separable" from the transformation and we have width $k_X \cdot W(\theta) = U(\theta) - L(\theta)$, which is proportionate to the width function in our initial analysis of HDRs for a unimodal density. In this case the optimal confidence interval problem reduces to computation of a HDR for the density under consideration (which may or may not be unimodal). Below we will look at an example of an optimal confidence interval problem where we have a separable transformation, where the optimisation can be framed using an initial strictly decreasing transformation $h_X$, or using an increasing linear function.

**EXAMPLE 1:** Consider the problem of finding an optimal (i.e., shortest) confidence interval for the true variance of a mesokurtic population, using the sample variance (see e.g., Tate and Klett 1959). This confidence interval is based on the well-known pivotal quantity:

$$\frac{S_n^2}{\sigma^2} \sim \text{Chisq}(n-1).$$

This distribution is exact for an underlying normal population, and it is an approximation based on the central limit theorem that is valid for any underlying mesokurtic population (see O'Neill 2016 for a generalisation that adjusts to allow for a non-mesokurtic population, and see Royall 1986 for some general theory regarding robust variance estimation). If we let $[L(\theta), U(\theta)]$ be an interval with coverage probability $1 - \alpha$ then we obtain the probability interval:

$$1 - \alpha = \mathbb{P}\big(L(\theta) \leq S_n^2/\sigma^2 \leq U(\theta)\big)$$
$$= \mathbb{P}(L(\theta)\sigma^2 \leq S_n^2 \leq U(\theta)\sigma^2)$$
$$= \mathbb{P}\left(\frac{S_n^2}{U(\theta)} \leq \sigma^2 \leq \frac{S_n^2}{L(\theta)}\right).$$

In this case we can see that the observable value $S_n^2$ is "separable" from the interval, meaning that the probability interval is proportionate in length to $S_n^2$. Thus, we may ignore this value and consider the confidence interval problem using the strictly decreasing continuous function $h(\sigma^2) = 1/\sigma^2$ with inverse function $g(\sigma^2) = 1/\sigma^2$. Thus, in this problem, the width function and its derivatives are given by:

$$W(\theta) = \frac{1}{L(\theta)} - \frac{1}{U(\theta)},$$

$$W'(\theta) = \frac{1}{f(U(\theta))} \cdot \frac{1}{U(\theta)^2} - \frac{1}{f(L(\theta))} \cdot \frac{1}{L(\theta)^2},$$

$$W''(\theta) = \frac{2/L(\theta) + u(L(\theta))}{f(L(\theta))^2} \cdot \frac{1}{L(\theta)^2} - \frac{2/U(\theta) + u(U(\theta))}{f(U(\theta))^2} \cdot \frac{1}{U(\theta)^2}.$$

Using the density function of the chi-squared distribution, it can be show that:

$$W''(\theta) = \text{pos. const} \times \left[(n + 1 - L) \cdot L^{n-6} \cdot \exp(L) - (n + 1 - U) \cdot U^{n-6} \cdot \exp(U)\right].$$

Over all values giving bounds $L(\theta) < n + 1 < U(\theta)$ the width function is strictly convex, so there is a unique critical point $\hat{\theta}$ that optimises the confidence interval. The convexity of the width function makes it easy to obtain the optimising value with numerical methods. □

**EXAMPLE 2:** The pivotal quantity in Example 1 can be reframed in its inverse form as:

$$\frac{\sigma^2}{S_n^2} \sim \text{InvGamma}\left(\frac{n-1}{2}, \frac{1}{2}\right).$$

This form makes it easier to write the confidence interval. If we let $[L(\theta), U(\theta)]$ be an interval with coverage probability $1 - \alpha$ then we obtain the probability interval:

$$1 - \alpha = \mathbb{P}(L(\theta) \leq \sigma^2/S_n^2 \leq U(\theta))$$
$$= \mathbb{P}(L(\theta)S_n^2 \leq \sigma^2 \leq U(\theta)S_n^2).$$

In this form the observable value $S_n^2$ is again "separable" from the interval, and we have the strictly increasing continuous function $h(\sigma^2) = 1$ with inverse function $g(\sigma^2) = 1$. Thus, in this problem, the width function is $W(\theta) = U(\theta) - L(\theta)$ and its derivatives are given by:

$$W'(\theta) = \frac{1}{f(U(\theta))} - \frac{1}{f(L(\theta))},$$

$$W''(\theta) = \frac{u(L(\theta))}{f(L(\theta))^2} - \frac{u(U(\theta))}{f(U(\theta))^2}.$$

Since the inverse-gamma density is strictly quasi-concave, we can restrict attention to the "admissible range" of parameters where the interval spans the mode (which is $1/(n+1)$ in this case) and this ensures that the width function is strictly convex. It is therefore simple to use numerical methods to obtain the minimising value $\hat{\theta}$, and thereby obtain the bounds of the optimal confidence interval. ☐

## 11. Comparison to other R packages — accuracy and speed

The HDR computation methods in this paper are programmed in the **stat.extend** package (O'Neill and Fultz 2020). This is done via functions of the form **HDR.xxxx** where the suffix specifies a distributional family for the computation using standard R syntax. These functions employ optimisation methods that depend on the shape of the density under the stipulated parameters, using underlying methods for monotone, unimodal and bimodal densities. There are three other R packages that can compute HDRs for known probability distributions: the **hdrcde** package (Hyndman, Einbeck and Wand 2018), the **pdqr** package (Chasnovski 2019) and the **HDInterval** package (Meredith and Kruschke 2019).[20] In this section we compare these packages over four example distributions. We will see that the **stat.extend** package is **more accurate and faster** than other packages for computations that are in its scope. The package also provides some additional user-friendly features including outputs in set format.[21] The only drawback of the **stat.extend** package compared to these other packages is that it is limited in scope, and only provides HDR functions for specific families of distributions; the other packages use more general methods that can be applied to any distribution.

---

[20] The **mclust** package (Fraley *et al.* 2020) computes the density cut-off but does not compute the HDR.
[21] The HDR functions in the **stat.extend** package use a "defensive programming" method which exhaustively checks all the inputs to the functions to ensure that they are of the correct type. The functions also converts the computed HDR to a set object (using the **sets** package) and makes other changes to the output to make it more user-friendly.

For our comparison we will compute HDRs for four examples, two that are unimodal and two that are bimodal. For each HDR we compute the disparity between the coverage probability and the stipulated minimum coverage probability (which we call the **probability disparity**) and the maximum disparity between density values at the endpoints of the region, other than endpoints that are boundaries of the support (which we call the **density disparity**). In a perfect HDR these disparities are zero, so the accuracy of the methods and package implementations can be observed by looking at the magnitude of these disparities. In addition to these two measures of accuracy we look at the number of intervals in the computed HDR (which can be compared to the true HDR) and whether or not the computed HDR includes any regions that are outside the support of the distribution. We compute these four indicators of accuracy for each of the packages operating on each of the four examples.

In addition to measuring the accuracy of the packages we also measure the computation speed using the `microbenchmark` package (Mersmann 2019) using one-thousand computations of each command. In order to reduce dependence on machine speed, we measure the "relative speed" of each computation, expressed as a multiplier of the time taken by `R` to sort the integers $1, \ldots, 10^6$ into descending order on the same machine.[22] Some of the packages that compute HDRs do so using the "exact" density and quantile functions, but some use alternative inputs, such as finite lists containing variable and density values, which are then employed using grid methods. In these cases, the user has a choice of the "fineness" of the input list, which yields a trade-off between accuracy and speed — a finer set of values gives a more accurate computation but takes longer. For all computations of this kind we have chosen to use a list of one-thousand equally spaced values for the density list, with the end-point of the list going either to the bound of the support (for bounded supports) or three standard deviations above the mean (for unbounded supports). Robustness testing by the author found that using a finer list of values gave only relatively small increases in accuracy compared to the additional computation time. For assessing the speed of these computations we include the time taken to create the density list as part of the computation. While we have done what we can to try to make the computations of speed "objective", our speed calculations are not replicable due to variations in machine speed and computational requirements of background processes.

---

[22] This computation is the function `sort(1L:1000000L, decreasing = TRUE)`, which uses functions in the base package.

| Package | Probability disparity | Density disparity | Ints | Outside support | Relative Speed† |
|---|---|---|---|---|---|
| **Chi-squared HDR (unimodal)** | | | | | |
| (Chi-squared distribution with df = 4 degrees-of-freedom and non-centrality parameter NCP = 2) | | | | | |
| `stat.extend` | $7.661 \times 10^{-15}$ | $9.468 \times 10^{-11}$ | 1 | No | $0.917 \times$ |
| `HDInterval` | $7.772 \times 10^{-15}$ | $8.490 \times 10^{-9}$ | 1 | No | $0.882 \times$ |
| `hdrcde`* | $1.215 \times 10^{-2}$ | $1.679 \times 10^{-6}$ | 1 | No | $0.254 \times$ |
| `pdqr` | $2.527 \times 10^{-3}$ | $1.757 \times 10^{-4}$ | 1 | No | $2.315 \times$ |
| **Gamma HDR (unimodal)** | | | | | |
| (Gamma distribution with as shape $s = 3$ and scale $\lambda = 4$) | | | | | |
| `stat.extend` | 0 | $1.753 \times 10^{-13}$ | 1 | No | $0.051 \times$ |
| `HDInterval` | 0 | $2.180 \times 10^{-10}$ | 1 | No | $0.021 \times$ |
| `hdrcde`* | $1.753 \times 10^{-2}$ | $6.156 \times 10^{-8}$ | 1 | No | $0.257 \times$ |
| `pdqr` | $4.177 \times 10^{-3}$ | $1.212 \times 10^{-6}$ | 1 | No | $1.631 \times$ |
| **Beta HDR (bimodal)** | | | | | |
| (Beta distribution has with parameters $s_1 = 0.40$ and $s_2 = 0.60$) | | | | | |
| `stat.extend` | $1.665 \times 10^{-16}$ | $1.110 \times 10^{-16}$ | 2 | No | $0.116 \times$ |
| `HDInterval`** | $1.162 \times 10^{-2}$ | $9.327 \times 10^{-3}$ | 3 ✘ | Yes ✘ | $29.290 \times$ |
| `hdrcde`** | $1.615 \times 10^{-3}$ | $2.465 \times 10^{-2}$ | 18 ✘ | Yes ✘ | $156.579 \times$ |
| `pdqr` | $4.313 \times 10^{-4}$ | $3.729 \times 10^{-9}$ | 2 | Yes ✘ | $9.481 \times$ |
| **Beta HDR (extreme bimodality)** | | | | | |
| (Beta distribution has with parameters $s_1 = 0.03$ and $s_2 = 0.05$) | | | | | |
| `stat.extend` | 0 | $4.728 \times 10^{-9}$ | 2 | No | $0.306 \times$ |
| `HDInterval`** | $1.165 \times 10^{-2}$ | $7.659 \times 10^{-3}$ | 2 | Yes ✘ | $30.172 \times$ |
| `hdrcde`** | $8.000 \times 10^{-1}$ | $2.372 \times 10^{-3}$ | 2 | No | $117.944 \times$ |
| `pdqr` | $9.992 \times 10^{-5}$ | $2.169 \times 10^{-7}$ | 2 | Yes ✘ | $53.477 \times$ |

† Relative speed is a multiplier compared to the time taken for a baseline computation; the baseline computation is sorting one-million ascending integers into descending order
* Computations depend on a finite list of density points using $10^3$ values; calculations for computation speed include generation of the density list.
** Computations depend on a finite vector of $10^6$ simulated pseudo-random values; calculations for computation speed include generation of the pseudo-random values.
✘ This outcome is not consistent with the true HDR

**Table 1:** Accuracy of HDRs computed by different `R` packages

In Table 1 we record information about the accuracy and speed for these packages. As seen from the results in the table, the accuracy of the `stat.extend` package compares favourably to other packages. All of the packages compute the HDR for the unimodal densities with reasonable accuracy, but `stat.extend` and `HDInterval` both give high accuracy. For the first bimodal density, all other packages compute a HDR that goes outside the support of the distribution, and two of them give HDRs that have the wrong number of intervals. The accuracy of some of these other packages is quite poor here, though this level of accuracy depends on the number of simulated values used in the functions, and so it can be improved with more simulations. For the more extreme bimodal density, two of the other packages compute a HDR that goes outside the support of the distribution, and the other package (the **hdrcde** package) gives a region that roughly "inverts" the true HDR (i.e., it gives a HDR that is roughly equal to the set difference of the support and the true HDR). Consequently, the probability disparity for this case is extremely large.[23]

The relative speeds of the packages depend a great deal on the particular examples used. For computations of HDRs for simple unimodal densities, the **stat.extend** package is slower than other packages, but for computations of HDRs for bimodal densities it is much faster than other packages. The main reason for this is that the computations in other packages become substantially more complex for bimodal densities, but the **stat.extend** package is able to solve these using an optimisation method that is similar to the unimodal case. If we take an average over all four examples then the **stat.extend** package is substantially faster than the other packages. (This holds even if we remove the time taken to compute density lists or simulated values from the **HDInterval** and **hdrcde** packages.)

**12. Discussion and conclusion**

In the above sections, we have used nonlinear optimisation to compute HDRs for continuous univariate distributions. We have shown how the problem of finding the HDR can be framed as an appropriate optimisation problem, and converted to an unconstrained problem. Using a set of general algorithms for densities with a given shape (monotone, quasi-concave, quasi-convex) we can build up individualised functions to compute HDRs for a distributional family, by taking in the input parameters and then applying the appropriate algorithm for the shape of the density.

The above methods have been used to program a suite of functions for each of the standard distributional families in base **R** and various distributions in extension packages. These are available in the **R** package **stat.extend** (O'Neill and Fultz 2020). This package builds on the **sets** package to compute HDR functions; the output shows the HDR, plus some additional information on the coverage probability and the optimisation method.[24] The package also gives

---

[23] Since the computed HDR roughly inverts the true HDR, its coverage probability is approximately $\alpha$, and so the probability disparity is approximately $1 - 2\alpha$.

[24] The code for the HDR functions in this package uses the algorithms in this paper, but there is also additional code to check inputs and add some further elements to the output. The underlying programming uses "helper"

functions to compute confidence interval functions, giving an output that is a set showing the interval, plus additional information on the optimisation method. The methods set out in this paper can be used to create similar HDR functions for other families of continuous univariate distributions, or confidence interval functions using other pivotal quantities, so long as it is possible for the programmer to determine the general shape of the distribution from the input parameterisation. (This may require the programmer to do some calculus to find the slope properties of the distribution they are working with and determine the appropriate shape for each possible value of the parameters.)

Our goal in this paper has been to plug a practical gap in the statistical literature to give clear instructions for computing HDRs for classes of univariate densities. We have been motivated primarily by the fact that statistical programming languages commonly include probability functions (density, quantile function, etc.) for the standard distributional families, but often do not include corresponding functions for computing HDRs from those distributional families. Similarly, such languages often lack functions to compute optimal confidence intervals for cases where the distribution under use is not symmetric. We hope that this paper is able to assist to ensure that statistical computing facilities include simple functions to compute HDRs for all standard classes of densities.

---

functions that split the coding into pieces. Code for functions used in the **stat.extend** package can be called in **R** using the relevant function commands (e.g., **HDR.monotone**, **HDR.unimodal**, **HDR.bimodal**, **HDR.discrete.unimodal**, **HDR.discrete**, etc.).

**Acknowledgement:** The author is grateful to two anonymous referees for their comments on an earlier draft of this paper. The comparison of accuracy and speed against other packages was added due to their excellent suggestions.

**Conflict of Interest Statement**

On behalf of all authors, the corresponding author states that there is no conflict of interest.

# Appendix: Proof of Theorems

In this appendix we give proofs of the lemmas and theorems shown in the main body of the paper. For ease of reference, we repeat these lemmas and theorems here.

**LEMMA 1:** If $\mathcal{H}$ is a highest density region then for any set $\mathcal{A}$ we have:

$$|\mathcal{A}| < |\mathcal{H}| \quad \Longrightarrow \quad \mathbb{P}(X \in \mathcal{A}) < \mathbb{P}(X \in \mathcal{H}),$$
$$|\mathcal{A}| \leq |\mathcal{H}| \quad \Longrightarrow \quad \mathbb{P}(X \in \mathcal{A}) \leq \mathbb{P}(X \in \mathcal{H}).$$

**PROOF OF LEMMA 1:** For any set $\mathcal{A}$ we have:

$$\mathbb{P}(X \in \mathcal{H}) - \mathbb{P}(X \in \mathcal{A}) = \int_{\mathcal{H}} f(x)dx - \int_{\mathcal{A}} f(x)dx$$
$$= \int_{\mathcal{H}-\mathcal{A}} f(x)dx - \int_{\mathcal{A}-\mathcal{H}} f(x)dx.$$

If $|\mathcal{A}| < |\mathcal{H}|$ then we must have $|\mathcal{A} - \mathcal{H}| < |\mathcal{H} - \mathcal{A}|$, so the range of integration in the first integral is larger than in the second integral. Moreover, since $\mathcal{H}$ is a highest-density region it encompasses all points with $f(x) \geq f_*$, so we must have $f(x) < f_* \leq f(y)$ for all $x \in \mathcal{A} - \mathcal{H}$ and $y \in \mathcal{H} - \mathcal{A}$. This means that the integrand in the first integral is strictly higher than in the second integral.[25] Since the first integral has a higher integrand and larger region of integration, this implies that it is larger than the second integral, so $\mathbb{P}(X \in \mathcal{H}) > \mathbb{P}(X \in \mathcal{A})$. This gives us the first inequality result in the lemma; the second result follows analogously. ∎

**THEOREM 1:** A highest density region $\mathcal{H}$ with actual coverage probability $1 - \alpha$ is a smallest closed covering region with minimal coverage probability $1 - \alpha$.

**PROOF OF THEOREM 1:** Consider a HDR $\mathcal{H}$ with actual coverage probability $1 - \alpha$. To show that $\mathcal{H}$ is a smallest closed covering region with minimal coverage probability $1 - \alpha$, we have to show that there is no closed set $\mathcal{A}$ with $|\mathcal{A}| < |\mathcal{H}|$ and $\mathbb{P}(X \in \mathcal{A}) \geq 1 - \alpha$. This follows directly from the first inequality result in Lemma 1. ∎

---

[25] One other possibility is that $\mathcal{A} \subset \mathcal{H}$ so that $\mathcal{A} - \mathcal{H} = \emptyset$ and so no density value exists over this set. In this case the second integral is zero and the first is positive, so the first integral is still strictly higher than the second.

**THEOREM 2:** If the intensity function for $X$ is continuous then a highest density region $\mathcal{H}$ formed with minimal coverage probability $1-\alpha$ has actual coverage probability $1-\alpha$.

**PROOF OF THEOREM 2:** Consider a HDR $\mathcal{H}$ with minimum coverage probability $1-\alpha$. Since the intensity function $H$ is continuous, the random variable $X$ is also continuous, so we have:
$$\mathbb{P}(X \in \mathcal{S}) = \mathbb{P}(X \in \text{closure}\,\mathcal{S}) \qquad \text{for any set } \mathcal{S}.$$
Since the intensity function $H$ is continuous we have $H(f_*) = 1-\alpha$, it follows that:
$$\begin{aligned}
\mathbb{P}(X \in \mathcal{H}) &= \mathbb{P}(X \in \text{closure}\{x \in \mathbb{R} | f(x) \geq f_*\}) \\
&= \mathbb{P}(X \in \{x \in \mathbb{R} | f(x) \geq f_*\}) \\
&= \mathbb{P}(f(X) \geq f_*) \\
&= H(f_*) = 1 - \alpha,
\end{aligned}$$
which establishes that the actual coverage probability is $1-\alpha$. ∎

---

**Algorithm 6: Print method for HDRs**

This function gives a customised print method for 'hdr' objects.

**Inputs:** An **object** of class 'hdr'
**Output:** User friendly print output for the HDR

```
print.hdr <- function(object) {

  #Print description of HDR
  cat('\n      Highest Density Region (HDR) \n \n');
  cat(paste0(sprintf(100*attributes(object)$probability, fmt = '%#.2f'), '%'),
      'HDR for', attributes(object)$distribution, '\n')

  #Print method
  if (!is.na(attributes(object)$method)) {
    cat(attributes(object)$method, '\n') }

  #Print HDR interval
  cat('\n')
  print(c(object))
  cat('\n') }
```